\documentstyle[sprocl,epsf]{article}
\def\be{\begin{equation}}
\def\ee{\end{equation}}
\def\bea{\begin{eqnarray}}
\def\eea{\end{eqnarray}}

\begin{document}
\title{Virtual Compton Scattering and Generalized Polarizabilities of
  the Nucleon in Heavy Baryon Chiral Perturbation
  Theory}

\author{Thomas R. Hemmert and Barry R. Holstein}

\address{Department of Physics and Astronomy\\
 University of Massachusetts, Amherst, MA 01003, USA}

\author{{Germar Kn\"ochlein}~\footnote{Talk given 
at the {\em Workshop on Virtual Compton
      Scattering}, Clermont-Ferrand, June 1996} and Stefan Scherer}

\address{Institut f\"ur Kernphysik, Becherweg 45\\
Johannes Gutenberg-Universit\"at, D-55099 Mainz, Germany}


\maketitle\abstracts{
The spin-independent part of the virtual Compton scattering (VCS)
amplitude from the nucleon is calculated within the framework of heavy
baryon chiral perturbation theory (HBChPT). The calculation is performed
to third order in external momenta according to chiral power
counting. The relation of the tree-level amplitudes to what is
expected from the
low-energy theorem is
discussed. We relate
the one-loop results to the structure coefficients of
a low-energy expansion for the model-dependent part of the VCS
amplitude recently defined by Fearing and Scherer. Finally we discuss
the connection of our results with the
generalized polarizabilities of the nucleon defined by
Guichon, Liu and Thomas.
}

\section{Introduction}

With the growing number of experimental
proposals~\cite{props} seeking to
investigate virtual Compton scattering (VCS) off a baryonic target there 
has been correspondingly increased activity within the theoretical community
examining VCS from the nucleon~\cite{GLT,Van,KKS1,GLT2,MD}
as well as from nuclei~\cite{nuclei}
in recent months. The
primary objective of the real Compton scattering experiments involving the
nucleon has traditionally been 
the determination of the nucleon electric and magnetic polarizabilities
--- $\alpha_0$ and $\beta_0$.\cite{exp} On the theoretical side these 
polarizabilities have been calculated within various models (see, {\it
  e.g.}, the review articles by Holstein~\cite{Holsteinp} and
L'vov~\cite{Lvovp}).
The pioneering work on
the generalization of the electromagnetic polarizabilities of the
nucleon to the VCS case~\cite{GLT} (here we only refer to the
spin-averaged amplitude),
\begin{equation}
\gamma^*(\varepsilon^{\mu},q^{\mu}) + N(p_i^{\mu}) \rightarrow
\gamma(\varepsilon'^{\mu},{q'}^{\mu}) +
N(p_f^{\mu})\,\,\,\,\,\,
({q'}^2=0\, , \,\,\,q^2=-Q^2<0 )\, ,
\label{process}
\end{equation}
made a first prediction for the generalized
electric and magnetic polarizability as a function of the initial
photon three-momentum, $\alpha \left( \mid \vec q \mid \right)$ and
$\beta \left( \mid \vec q \mid \right)$, within the framework of a
non-relativistic constituent quark model~\cite{GLT}. Subsequently, an
effective Lagrangian approach~\cite{Van},
a refined version of the quark model~\cite{GLT2}
and the linear sigma model~\cite{MD}
have been used to make predictions for the
generalized polarizabilities. Herein we will add the prediction of chiral
perturbation theory (ChPT),
the effective description of quantum
chromodynamics in the nonperturbative energy regime with nucleons and
pions as physical degrees of freedom~\cite{CPTintro,BKKM,Ecker},
to third order in the external
momenta --- $O \left( p^3 \right)$,
generalizing the work of
Bernard {et al.}~\cite{BKKM,BKMS}
on the real Compton process to VCS. In order to
achieve this, it is necessary to decompose the invariant VCS
amplitude ${\cal{M}}_{VCS}$ into a structure-dependent part, which
is regular and consists of contributions from one-particle irreducible Feynman
diagrams (for the spin-independent part of the VCS amplitude these are
the same loop diagrams as in Bernard {et al.}~\cite{BKKM}),
and a singular structure- or model-independent part,
which is generated
by  $s$- and $u$-channel Born diagrams, together with the seagull graph
required by gauge invariance. For this decomposition we follow the
convention of Guichon {et al.}~\cite{GLT} and Scherer
{et al.}~\cite{KKS1}, where the Born contribution
is calculated in terms of Dirac and
Pauli form factors, $F_1$ and $F_2$. Recently, a general
low-energy parametrization of the structure-dependent
part of the VCS amplitude on a spin 0 target has been developed by
Fearing and Scherer~\cite{FS1}. Since it has been shown by Bernab\'eu
and Tarrach~\cite{BT} that the form of the {\it spin-independent}
part of the VCS amplitude for a spin $\frac{1}{2}$
target is the same as for a spin $0$ target, we can apply this
parametrization to our calculation in order to determine the nine structure
coefficients in the low-energy expansion to fourth order in the
four-momenta of the initial and final state photons. Finally we will
relate these structure coefficients to the generalized
polarizabilities of the multipole expansion by Guichon
{et al.}~\cite{GLT}.
A more comprehensive discussion of the
material presented in this contribution can be found in 
Hemmert {et al.}~\cite{HHKS}.

\section{Chiral Calculation of the spin-independent VCS Amplitudes}
The invariant amplitude for VCS can be decomposed
into a
transverse and a longitudinal part and,
using current conservation,
can be written as~\cite{KKS1}
\begin{equation}
{\cal{M}}_{VCS} = - i e^2 \varepsilon_{\mu} M^{\mu} \, = - i e^2
\varepsilon_{\mu} \varepsilon'^*_{\nu} M^{\mu \nu}
= i e^2 \left( \vec \varepsilon_T \cdot \vec M_T +
  \frac{q^2}{\omega^2} \varepsilon_z M_z \right)
\, .
\label{invamp}
\end{equation}
Since we will restrict our discussion to the spin-averaged component
of 
${\cal{M}}_{VCS}$ in the following (the analogous spin-dependent
calculation will be the subject of a future publication),
the invariant amplitude can be
parametrized in terms of three independent structures,
\begin{equation}
{\cal{M}}_{VCS}^{non-spin}
= i e^2 \left[
\vec \varepsilon\,'^*
\cdot \vec \varepsilon_T A_1 + \vec \varepsilon\,'^* \cdot {\hat{q}}
\vec \varepsilon_T \cdot {\hat{q}}' A_2 + \vec \varepsilon\,'^*
\cdot {\hat{q}}
\frac{q^2}{\omega^2} \varepsilon_z
A_9 \right] \, ,
\end{equation}
two of which ($A_1$ and $A_2$) are purely transverse and one ($A_9$) which is
longitudinal.
A standard calculation of the $s$- and
$u$-channel Born terms together with the seagull contribution with the
Lagrangian of HBChPT (see, {\it
  e.g.}, Bernard {et al.}~\cite{BKKM} and Ecker and 
Moj\v{z}i\v{s}~\cite{Ecker})
yields the tree-level results
\begin{eqnarray}
A_1^{tree} & = & - \frac{1}{2 M} \left( 1 + \tau_3 \right)\,
,
\nonumber\\
A_2^{tree} & = & \frac{1}{2 M^2} \left( 1 + \tau_3 \right)
\mid \vec q \mid \, ,
\nonumber\\
A_9^{tree} & = & \left( 1 + \tau_3 \right) \left(
- \frac{1}{2 M} +
\frac{1}{2 M^2} \mid \vec q \mid \cos \theta
+ \frac{1}{4 M^2} \frac{\mid \vec q
  \mid^2}{\omega'} \right)
\, ,
\end{eqnarray}
where $M$ is the nucleon mass and $\tau_3$ the isospin operator.
We have written the results in terms of three independent kinematical
quantities in the c.m.\ system --- the energy of 
the outgoing photon $\omega'$, the three-momentum of the
initial state photon $|\vec q|$ and the angle between the
three-momenta of the initial and final state photons $\theta$. We
have used approximations for the initial photon
energy $\omega$ and the Lorentz invariant quantities $t = \left( q -
  q' \right)^2$ and $q^2$,
\begin{eqnarray}
\omega & = & \omega' + O\left(r^2/M\right) \, , \label{ompr}\nonumber\\
t & = &  - \omega'^2 - \mid \vec q \mid^2 + 2 \omega' \mid \vec q \mid
\cos \theta + {{O}}\left(r^3/M\right) \, , \label{t}\nonumber\\
q^2 & = & \omega'^2 - \mid \vec q \mid^2 +
{{O}}\left(r^3/M \right) \,\,\,\,\,\,\, (r \in \{ \omega', \mid \vec
q \mid \}), \label{q2}
\end{eqnarray}
in order to be consistent with chiral power counting at $O
\left( p^3 \right)$, the chiral order to which our calculation applies. The
structure-dependent part of the VCS amplitude is generated by nine 
diagrams~\cite{BKKM}. The evaluation of the diagrams is
straightforward but tedious. Using Eq. (\ref{ompr}) 
and expanding the loop integrals in $r/m_{\pi}$, we obtain to order $r^4$:\
\footnote{Here $r$ is a generic small kinematic quantity such as 
$
\omega',|\vec{q}|$.}
\begin{eqnarray}
A_1^{loop} & = & \frac{g_A^2}{F^2} \frac{1}{\pi m_{\pi}} \left[
\frac{5}{96} \omega'^2 + \frac{1}{192}
\omega'
\mid \vec q \mid \cos \theta \right. \nonumber \\
& &
 + \frac{17}{1920} \frac{1}{m_{\pi}^2} \omega'^4 +
 \frac{19}{1920} \frac{1}{m_{\pi}^2} \omega'^3 \mid \vec
 q \mid \cos \theta
 - \frac{1}{384} \frac{1}{m_{\pi}^2} \omega'^2 \mid \vec q
 \mid^2 \nonumber \\
& & - \frac{1}{320} \frac{1}{m_{\pi}^2}
 \omega'^2 \mid \vec q \mid^2 \cos^2 \theta
 \left. + \frac{1}{960} \frac{1}{m_{\pi}^2} \omega' \mid \vec
   q \mid^3 \cos \theta
 \right] \, , \label{a1l}
\\
A_2^{loop} & = & \frac{g_A^2}{F^2}
\frac{1}{\pi m_{\pi}}
\left[ -\frac{1}{192} \omega' \mid \vec q \mid - \frac{1}{384}
  \frac{1}{m_{\pi}^2} \omega'^3 \mid \vec q \mid\right. \nonumber\\
&&+
\left.\frac{1}{320} \frac{1}{m_{\pi}^2} \omega'^2 \mid \vec q \mid^2 \cos
\theta
- \frac{1}{960}
\frac{1}{ m_{\pi}^2} \omega' \mid \vec q \mid^3 \right] \, ,
\label{a2l}
\\
A_9^{loop} & = & \frac{g_A^2}{F^2} \frac{1}{\pi m_{\pi}}
\left[ \frac{5}{96} \omega'^2
  + \frac{17}{1920} \frac{1}{m_{\pi}^2} \omega'^4
+ \frac{7}{960} \frac{1}{m_{\pi}^2} \omega'^3 \mid \vec q
  \mid \cos \theta\right. \nonumber\\
&&- \left.\frac{7}{960} \frac{1}{m_{\pi}^2}
\omega'^2 \mid \vec q \mid^2 \right]
\, . \label{a9l}
\end{eqnarray}
Note that the loop results for VCS from a proton or a neutron are
identical, whereas the tree-level component vanishes for the neutron, because
to $O
\left( p^3 \right)$
in the spin-averaged part of the amplitude only the photon coupling to
the electric
charge and not to the magnetic moment generates non-zero contributions.

\section{Structure Coefficients and Generalized Polarizabilities}
In Fearing and Scherer~\cite{FS1} a general low-energy
parametrization of the structure-dependent
VCS amplitude on a spin $0$ target based on Lorentz and gauge
invariance
and the discrete symmetries has been generated to fourth
order in the photon four-momenta. Using the results of Bernab\'eu and
Tarrach~\cite{BT} such a low-energy expansion also applies
to the spin-averaged part of the VCS amplitude from a spin
$\frac{1}{2}$ target. We have determined the {\it a priori} unknown
structure constants in this expansion,
\begin{eqnarray}
g_0 = \frac{1}{192} \frac{g_A^2}{F^2} \frac{1}{\pi m_{\pi}}
\, ,
\hphantom{88888888_.}
&&
{\tilde{c}}_1 = - \frac{11}{192} \frac{1}{8 M^2} \frac{
g_A^2}{F^2}
\frac{1}{\pi m_{\pi}}\, ,
\nonumber\\
g_{2a} = \frac{1}{320} \frac{g_A^2}{F^2} \frac{1}{\pi
m_{\pi}^3}\, ,
\hphantom{8888888.}
&&
g_{2b} = - \frac{1}{960} \frac{g_A^2}{F^2} \frac{1}{\pi m_{\pi}^3}
\, ,
\nonumber\\
g_{2c} = \frac{1}{1920} \frac{1}{16 M^2}
\frac{g_A^2}{F^2} \frac{1}{\pi m_{\pi}^3}\, ,
\hphantom{..}
&&
c_3 = - \frac{3}{1280} \frac{1}{8 M^2}
\frac{g_A^2}{F^2} \frac{1}{\pi m_{\pi}^3}\, ,
\nonumber\\
{\tilde{c}}_{3a} = \frac{1}{240} \frac{1}{8 M^2}
\frac{g_A^2}{F^2} \frac{1}{\pi m_{\pi}^3}\, ,
\hphantom{881}
&&
{\tilde{c}}_{3b} = - \frac{1}{256} \frac{1}{8 M^2}
\frac{g_A^2}{F^2} \frac{1}{\pi m_{\pi}^3}\, , \nonumber\\
{\tilde{c}}_{3c} =
-
\frac{9}{640} \frac{1}{128 M^4} \frac{g_A^2}{F^2}
\frac{1}{\pi m_{\pi}^3}\, ,&&
\end{eqnarray}
where the notation is the same as in Fearing and
Scherer~\cite{FS1}. Since our expansion of kinematical quantities,
which is adequate for the application of 
HBChPT at $O(p^3)$ is different from the multipole
expansion~\cite{GLT},
it is not obvious how to relate the two
approaches. However, starting from the general structure
analysis~\cite{FS1}  
and then applying Eq.\ (\ref{ompr}), 
one finds 
relations between 
$\alpha \left( \mid \vec q \mid \right)$ and
$\beta \left( \mid \vec q \mid \right)$, the
  generalized polarizabilities~\cite{GLT}
$P^{(01,01)0} \left( \mid \vec q \mid \right)$ and
$P^{(11,11)0} \left( \mid \vec q \mid \right)$, and the 
structure coefficients of the low-energy expansion~\cite{FS1}:
\begin{eqnarray}
\alpha \left( \mid \vec q \mid \right) & = & \alpha_0 \left( 1 -
  \frac{7}{50} \frac{\mid \vec q \mid^2}{m_{\pi}^2} + {{O}} \left(
    \frac{\mid \vec q \mid^4}{m_{\pi}^4} \right)  \right)
= - \frac{e^2}{4 \pi} \sqrt{\frac{3}{2}}
P^{(01,01)0} \left( \mid \vec q \mid \right)
\nonumber \\
& = &
- \frac{e^2}{4 \pi} ( g_0 + 8 M^2 {\tilde{c}}_1
- \mid \! \vec q
  \! \mid^2 ( g_{2b} + 8 M^2 ( c_3 + {\tilde{c}}_{3b} ) )
\nonumber \\
& & 
+ {{O}} (
(\mid \! \vec q \mid /m_{\pi})^4
)
)
\, ,
\label{alphaex} \\
\beta \left( \mid \! \vec q \! \mid \right) & = & \beta_0 \left( 1 +
  \frac{1}{5} \frac{\mid \! \vec q \! \mid^2}{m_{\pi}^2} + {{O}} \left(
    \frac{\mid \! \vec q \! \mid^4}{m_{\pi}^4} \right) \right)
= - \frac{e^2}{4 \pi} \sqrt{\frac{3}{8}}
P^{(11,11)0} \left( \mid \! \vec q \! \mid \right)
\nonumber \\
& = & \frac{e^2}{4 \pi} ( g_0 - \mid \! \vec q \! \mid^2 g_{2b}
+ {{O}} (
(\mid \! \vec q \mid/m_{\pi})^4
)
)
\, . \label{betaex}
\end{eqnarray}
These results agree with the corresponding terms of the 
linear sigma model calculation~\cite{MD}. 
Moreover, it is worth mentioning that it is not
possible to determine $\alpha$ from the transverse amplitude $A_1$,
because the way in which we expand kinematical quantities introduces
contributions from higher multipoles (see, e.g.\, the $\omega'^2 \mid
\vec q \mid^2 \cos^2 \theta$
term in Eq.\ (\ref{a1l})) and thus obscures the $L'=1$ part of the
amplitude
where we would expect $\alpha\left( \mid \vec q \mid \right) $.
However, in the longitudinal amplitude
($A_9$) terms of higher multipolarity are absent, which allows us to
extract $\alpha\left( \mid \vec q \mid \right) $. The third
scalar polarizability in the multipole
expansion~\cite{GLT}, ${\hat{P}}^{(01,1)0}$,
is $1/M$ suppressed and thus vanishes in our $O \left(
  p^3 \right)$ calculation.
However, a general, model-independent
analysis of the spin-averaged part of the VCS
amplitude~\cite{MDKS} confirms the result of the sigma model
calculation~\cite{MD} that this polarizability can in fact be written as 
a linear combination of
$\alpha$ and $\beta$. In comparison with the other
calculations~\cite{GLT,Van,GLT2} our value for the electric
polarizability of the proton, $\alpha_0 = 12.8 \times 10^{-4}
\mathrm{fm^3}$, is significantly larger, and for the magnetic
polarizability of the proton our result, $\beta_0 = 1.3 \times 10^{-4}
\mathrm{fm}^3$, is smaller. The absolute values of the slopes 
$\frac{d}{d \mid \vec q \mid^2} \alpha \left(
\mid \vec q \mid = 0
\right)$ and
$\frac{d}{d \mid \vec q \mid^2} \beta \left( \mid \vec q \mid = 0
\right)$ are found to be considerably larger than in the other
calculations, the slope of the magnetic polarizability even has a
different sign. An experiment where both $\omega'$ and $\mid \vec q
\mid$ are smaller than the pion mass should be able to
distinguish between the different theoretical predictions. In this
kinematical region the multipole expansion~\cite{GLT} to first order
in the final state photon energy may not reliably
parametrize the VCS amplitude, because, as has been shown above,
contributions from higher multipoles enter the amplitudes at the same
level as the slopes of the electromagnetic polarizabilities and, thus,
must not be neglected. Our
parametrization and calculation should be applied to this kinematical
region but not to an experiment with
arbitrary, but not necessarily small $\mid \vec q \mid$, which is the
main application of the multipole expansion~\cite{GLT}.

\section*{Acknowledgments}
The work of G.K. and S.S. has been supported by Deutsche
Forschungsgemeinschaft (SFB 201) and Studienstiftung des Deutschen
Volkes. The research of T.R.H. and B.R.H. is supported in part by the
National Science Foundation. G. K. is indebted to
the High Energy Theory and the Nuclear Physics Group at UMass
in Amherst for their kind hospitality. Moreover, G. K. and S. S. want to
thank Prof. D. Drechsel and A. Metz for many valuable
discussions.

\end{document}